\documentclass[letterpaper,11pt]{article}

\usepackage{jcappub} 

\usepackage{tcolorbox}
\usepackage{latexsym}
\usepackage{amsfonts}
\usepackage{mathrsfs}
\usepackage{amssymb,amscd}
\usepackage[all]{xy}
\usepackage{amsmath}
\usepackage{amsthm}
\usepackage{fancyhdr}
\usepackage{fancybox}
\usepackage{xcolor}
\usepackage{hyperref}
\usepackage{verbatim}
\usepackage{makeidx}
\usepackage{indentfirst}
\usepackage[english]{babel}
\usepackage{graphicx}
\usepackage{listings}
\usepackage{subfigure}
\usepackage{multirow}
\usepackage{lscape}
\usepackage{physics}
\usepackage{amsmath}
\usepackage{framed}
\usepackage{bbold}
\usepackage{siunitx}
\usepackage{comment}
\usepackage{cases}
\usepackage{booktabs}
\usepackage{diagbox}

\usepackage{bm}
\numberwithin{equation}{section}

\usepackage{xparse}

\usepackage{hyperref}
\hypersetup{
    colorlinks=true,
    linkcolor=blue,
    filecolor=magenta,      
    urlcolor=cyan,
    pdftitle={BOSS Skew Spectra},
    pdfpagemode=FullScreen,
    }

\newcommand{\Lagr}{\mathcal{L}}

\newcommand{\VEV}[1]{\left\langle #1 \right\rangle}

\newcommand{\kvec}{\bm{k}}
\newcommand{\qvec}{\bm{q}}
\newcommand{\knl}{k_{\rm NL}}
\def\skew{\tilde{P}}

\def\d{{\rm d}}

\def\k{{\boldsymbol k}}

\def\q{{\boldsymbol q}}

\def\r{{\boldsymbol r}}

\def\0{{\boldsymbol{0}}}

\def\G{{\cal G}}

\def\O{{\cal O}}

\def\fnl{f_{\rm NL}}

\newcommand{\fnle}{\fnl^{\rm equil}}
\newcommand{\fnlo}{\fnl^{\rm ortho}}
\newcommand{\tildeP}{\tilde{P}}

\def\beq{\begin{equation}}
\def\eeq{\end{equation}}
\def\be{\begin{equation}}
\def\ee{\end{equation}}

\title{Analysis of BOSS Galaxy Data with Weighted Skew-Spectra}
\author{Shu-Fan Chen$^{1}$, Priyesh Chakraborty$^{2}$, and Cora Dvorkin$^{3}$} 
\affiliation{Department of Physics, Harvard University, 17 Oxford Street,
  Cambridge, MA 02138, USA}
\emailAdd{$^1$shufan\_chen@g.harvard.edu, $^2$pchakraborty@g.harvard.edu,
  $^3$cdvorkin@g.harvard.edu,
 } 

  \abstract{We present the first application of the weighted skew-spectra to analyze non-Gaussian information in galaxy survey data. 
 Using the tree-level galaxy skew-spectra together with the one-loop 
 power spectrum multipoles, we analyze the Sloan Digital Sky Survey (SDSS)-III Baryon Oscillation Spectroscopic Survey (BOSS) galaxy clustering data, and target our search towards the equilateral bispectrum shape of primordial non-Gaussianity. We use the Effective Field Theory model for the galaxy power spectrum and bispectrum, and account for systematic effects, such as the survey geometry. From our likelihood analysis, we find $\fnle=-34^{+296}_{-334}$ at $68\%$ CL, consistent with previous works, while systematic errors from our treatment of the survey geometry lead to an unreliable estimation of $\fnlo$. We further constrain the bias and counterterm parameters, while keeping the cosmology fixed to $\textit{Planck }2018$ values. As a check, we also validate our analysis pipeline using the {\tt Nseries} simulation suite.}

\begin{document}
\maketitle
\flushbottom

\section{Introduction}

We have learned from precision cosmological measurements that our model of the universe at large scales can be described by a few parameters. However, the physics of the early universe remains to be fully elucidated. One interesting observable that can be used to learn about this period is primordial non-Gaussianity (PNG), which is a weak deviation from a pure Gaussian probability distribution function of matter and radiation fluctuations in the early universe. Such a signal could be generated by interactions during an early period of inflation, and therefore a detection of PNG could additionally provide a window into probing physics at unprecedented energy scales \cite{Bartolo:2004if,Komatsu:2009kd,Meerburg:2019qqi,Achucarro:2022qrl}. 

The simplest model of inflation, featuring a single field, can be captured by the so called Effective Field Theory (EFT) of inflation \cite{Cheung:2007st}, which produces two dominant types of primordial bispectra. One is the equilateral shape, which peaks when the wavenumbers of the fluctuations, $k_1$, $k_2$, and $k_3$, satisfy $k_1\sim k_2 \sim k_3$, and the other one is the orthogonal shape which peaks in the flattened configuration ($k_1\sim k_2\sim k_3/2$). Their amplitudes are parametrized by $\fnl^{\rm equil}$ and $\fnl^{\rm ortho}$, respectively. 

The leading constraints on $\fnl^{\rm equil}$ and $\fnl^{\rm ortho}$ have thus far been obtained through cosmic microwave background (CMB) observations. The \textit{Planck} 2018 satellite provided us with measurements of $\fnl^{\rm equil}=-26\pm 47$ and $\fnl^{\rm ortho}=-38\pm 24$ at $68\%$ CL \cite{Planck:2019kim}. It is expected that CMB constraints will be improved with the advent of upcoming large-scale structure (LSS) surveys, which in principle contain orders of magnitude more modes of information \cite{Desjacques:2010jw}. The search for PNG in the Sloan Digital Sky Survey (SDSS)-III Baryon Oscillation Spectroscopic Survey (BOSS) catalogue \cite{BOSS:2012dmf,BOSS:2016wmc} has already begun \cite{Cabass:2022wjy,Cabass:2022ymb,DAmico:2022gki,Leistedt:2014zqa,Castorina:2019wmr,Mueller:2021tqa}, and while the status is not currently at a competitive level with the CMB, these constraints are expected to improve with upcoming galaxy surveys such as DESI \cite{DESI:2016fyo}, the Vera Rubin Observatory \cite{2009arXiv0912.0201L}, SPHEREx \cite{Dore:2014cca}, and EUCLID \cite{Amendola:2016saw}, among others. 

A standard route for constraining PNG has been by using the galaxy bispectrum. A challenge with this approach is that estimating this statistic from a survey or mock is computationally expensive. Since the covariance matrix is typically needed to be estimated from many ($\O(1000)$) mock catalogs, performing a Markov Chain Monte Carlo (MCMC) likelihood analysis can prove to be a time-consuming task. To circumvent this issue, the bispectrum weighted skew-spectrum (henceforth skew-spectrum) was introduced as a tool in LSS analyses as a compression of the bispectrum \cite{Schmittfull:2014tca}. The skew-spectrum is obtained by averaging over the bispectrum using an appropriate kernel, so that the process of estimating it from data reduces to factorizable convolutions, which can all be reduced to a sequence of Fast Fourier Transforms (FFTs). This type of method was initially developed in order to expedite estimators of $\fnl$ from CMB temperature and polarization data \cite{Komatsu:2003iq}, and has been subsequently developed for the large-scale structure of the universe. Refs. \cite{Schmittfull:2014tca,MoradinezhadDizgah:2019xun} studied such skew-spectra for the galaxy bispectrum in Fourier space, and these were subsequently extended to redshift-space bispectra \cite{Schmittfull:2020hoi, Hou:2022rcd} and harmonic space for weak-lensing analyses \cite{Munshi:2020ofi,Chakraborty:2022aok}. In this paper, we present the first application of skew-spectra to LSS data. 

Using the BOSS DR12 galaxy catalogue \cite{BOSS:2012dmf,BOSS:2016wmc}, we focus on investigating PNG due to single-field inflation using the one-loop galaxy power spectrum multipoles and the tree-level galaxy skew-spectra. There are some differences in the precise redshift-space distortions (RSD) information used in the literature and the one we employ here. Ref. \cite{Cabass:2022wjy} makes use of the bispectrum monopole at tree-level and uses the window-free estimator \cite{Planck:2019kim} in order to de-contaminate the measured bispectrum from survey geometry effects. Ref. \cite{DAmico:2022gki} goes to 1-loop order in the bispectrum monopole and, instead, chooses to approximate window function effects using a phenomenological approximation \cite{Gil-Marin:2014sta}. In this work, we will use skew-spectra with multiple kernels to capture RSD information in the bispectrum, and we will apply a phenomenological approximation to the window effects.

The outline of this work is as follows: we first present the theoretical specifications of our model for the power and skew-spectra in \S \ref{sec: theory}, which uses the Effective Field Theory of large-scale structure framework. The estimation of skew-spectra from the BOSS galaxy catalogues and from simulations is discussed in \S \ref{sec: data}, and details regarding our inference methodology are provided in \S \ref{sec: inference}. Finally, we discuss our results in \S \ref{sec: results}. 

\section{The Theory Model}\label{sec: theory}

One of the main advantages of the skew-spectra is to optimally measure the amplitude parameters in the galaxy bispectrum, while being computationally much more efficient than the standard bispectrum. The amplitude $A_B$ of the galaxy bispectrum  $B_g(\kvec_1, \kvec_2, \kvec_3)$ is obtained through a minimum-variance estimator \cite{Fergusson:2010ia,Schmittfull:2014tca},
\begin{eqnarray}
\hat{A}_B &=& \int\frac{\d^3k}{N_{\rm th}}\int_{\qvec} \frac{B^{\rm th}_g(\kvec,\qvec,-\kvec-\qvec)}{P_{gg}(k)P_{gg}(q)P_{gg}(\abs{\kvec+\qvec})}\nonumber \\
&&\times\left[\delta_g(\kvec)\delta_g(\qvec)\delta_g(-\kvec-\qvec)-3\VEV{\delta_g(\kvec)\delta_g(\qvec)}\delta_g(-\kvec-\qvec)\right],
\end{eqnarray}
where $B^{\rm th}_g(\kvec_1,\kvec_2,\kvec_3)$ is the theoretical template for the galaxy bispectrum, $P_{gg}(k)$ is the galaxy power spectrum, and $N_{\rm th}$ is the number of $k$-modes. Hereafter, we will use the following notation: $\int_{\qvec}\equiv \int \frac{\d^3q}{(2\pi)^3}$. 

The amplitude estimator is a sum of the form $\hat{A}_B=\hat{A}_{B}^{(1)}+\hat{A}_{B}^{(3)}$, where $\hat{A}_{B}^{(1)}$ is a linear term in $\delta_g(\kvec)$ and $\hat{A}_{B}^{(3)}$ is cubic in $\delta_g(\kvec)$ \cite{Creminelli:2005hu}. We will ignore the linear term, which only has the effect of increasing the variance. To reduce clutter, we will still call the amplitude estimator $\hat{A}_{B}$ (instead of $\hat{A}_{B}^{(3)}$).

If the theoretical bispectrum can be written in a separable fashion, i.e. $B^{\rm th}_g(\kvec_1,\kvec_2, \kvec_3) = f(\kvec_1)g(\kvec_2) h(\kvec_3)$, the above estimator can be rewritten as:
\begin{eqnarray}
    \hat{A}_B &=& \int\frac{\d^3k}{N_{\rm th}} \left[\frac{f\delta_g}{P_{gg}}\star \frac{g\delta_g}{P_{gg}}\right](\kvec) \frac{h\delta_g}{P_{gg}}(-\kvec) \nonumber \\
    &=&4\pi\int\frac{\d k \,k^2}{N_{\rm th}}\hat{\skew}(k),
\end{eqnarray}
where $\hat{\skew}(k)$ is the skew-spectrum defined as:
\begin{eqnarray}\label{eq:skew_estimator}
    \hat{\skew}(k) &=& \int\frac{\d\hat{\kvec}}{4\pi}\left[\frac{f\delta_g}{P_{gg}}\star \frac{g\delta_g}{P_{gg}}\right](\kvec) \frac{h\delta_g}{P_{gg}}(-\kvec).
\end{eqnarray}

We can relax the optimality of this statistic and, instead of $B^{\rm th}_g$, choose any kernel $\mathcal{S}(\kvec,\kvec')$ to convolve with, as long as the kernel is still product separable. To be precise, given an arbitrary kernel $\mathcal{S}(\kvec,\kvec')=f(\kvec)g(\kvec')h(-\kvec-\kvec')$, the estimator is defined as,
\begin{eqnarray}\label{eq:skew_estimator}
    \hat{\skew}_{\mathcal{S}}(k) &=& \int\frac{\d\hat{\kvec}}{4\pi}\left[f\delta_g\star g\delta_g\right](\kvec) h\delta_g(-\kvec).
\end{eqnarray}
We will specify our choices for these kernels in \S\ref{sec:skewspec_kernels}. Note that given some arbitrary separable kernel $\mathcal{S}$, we can write the theory skew-spectrum as:
\begin{eqnarray}
    \tildeP_{\mathcal{S}}(k) &=& \VEV{\hat{\tildeP}_{\mathcal{S}}(k)} \nonumber \\
    &=& \int\frac{\d\hat{\kvec}}{4\pi}\int_{\qvec} \mathcal{S}(\qvec,\kvec-\qvec)B_g(\qvec,\kvec-\qvec,-\kvec),
\end{eqnarray}
where $B_g$ is the theoretical template for the galaxy bispectrum.

Therefore, in order to compute the skew-spectrum, it is necessary to first specify a theoretical model for the power spectrum and bispectrum. We will begin with a discussion of the primordial bispectrum shapes in \S\ref{sec:primordial}, and then lay out the details of the late-time galaxy power spectrum and bispectrum, under the EFT of LSS in \S\ref{sec:late time spectra}.

\subsection{Primordial Bispectrum}\label{sec:primordial}

By treating the inflaton field as a Goldstone mode of broken time-diffeomorphisms, the EFT of inflation generates interactions that are fixed by a few unknown speed-of-sound parameters \cite{Cheung:2007st, Cabass:2022wjy}. In this minimal scenario, the two cubic self-interactions of the inflaton field lead to two distinct shapes of bispectra, namely the equilateral and orthogonal shapes, with amplitudes $\fnle$ and $\fnlo$, respectively. The precise expression for the bispectrum of the primordial curvature perturbation $\zeta$ is given by \cite{Babich:2004gb,Senatore:2009gt}:
\begin{align}
    B_{\zeta}(k_1,k_2,k_3) = \frac{18}{5}\Delta_{\zeta}^4\frac{\mathcal{B}(k_1,k_2,k_3)}{k_1^2k_2^2k_3^2}.
\end{align}
Here $\Delta_\zeta^2\approx 4.1\times10^{-8}$ \cite{Planck:2018jri} is the amplitude of the primordial power spectrum, i.e. $k^3 P_\zeta(k)=\Delta_\zeta^2 (k/k_*)^{n_s-1}$, where $n_s$ is the spectral tilt and the pivot scale $k_*=0.05$ Mpc${}^{-1}$. The equilateral and orthogonal templates fix the shape function $\mathcal{B}$, given by:
\begin{align}
    \mathcal{B}_{\rm equil}(k_1,k_2,k_3) &= \left(\frac{k_1}{k_2}+5\text{ perms.}\right) - \left(\frac{k_1^2}{k_2k_3}+2\text{ perms.}\right) - 2\,,\\
    \mathcal{B}_{\rm ortho}(k_1,k_2,k_3) &= (1+p)\frac{\Delta}{e_3} - p\frac{\Gamma^3}{e_3^2}\,.
\end{align}
The various parameters appearing above are defined as:
\begin{align}
    &p = 8.52587\,, \qquad \Delta = (-k_1+k_2+k_3)(k_1-k_2+k_3)(k_1+k_2-k_3)\\
    &e_2 = k_1k_2 + k_2k_3 + k_1k_3\,, \qquad e_3=k_1k_2k_3\\
    &\Gamma = \frac{2}{3}e_2 - \frac{1}{3}(k_1^2+k_2^2+k_3^2)\,.
\end{align}
In the subsection below, we will discuss how these primordial shapes enter the late-time galaxy correlators.

\subsection{Late-Time Power Spectrum and Bispectrum}\label{sec:late time spectra}

The EFT of LSS is a perturbative model built to capture the gravitational evolution of galaxy correlators up to some nonlinear scale. It identifies an ultraviolet scale and provides a systematic method of including counterterms in order to appropriately renormalize the final correlators \cite{Carrasco:2012cv,Baumann:2010tm}. 

 The power spectrum and bispectrum of the density field $\delta_{L}$ in linear theory are given by:
\begin{eqnarray}\label{E:eftoflss_pt}
    \VEV{\delta_L(z,\kvec)\delta_L(z,\kvec')} &=& (2\pi)^3 \delta_D(\kvec+\kvec')P_L(z,k) \\
    \VEV{\delta_L(z,\kvec_1)\delta_L(z,\kvec_2)\delta_L(z,\kvec_3)} &=& (2\pi)^3 \delta_D(\kvec_{123})B_L(z;\kvec_1, \kvec_2, \kvec_3),
\end{eqnarray}
where the bispectrum $B_L$ is determined by the primordial bispectrum $B_{\zeta}$ as: 
\begin{eqnarray}
    B_L(z;\kvec_1, \kvec_2, \kvec_3) &=& \fnl T(z,k_1)T(z,k_2)T(z,k_3) B_{\zeta}(k_1, k_2, k_3).
\end{eqnarray}
Here, $\kvec_{123}\equiv \kvec_1+\kvec_2+\kvec_3$ (this notation will be followed henceforth) and $T(z,k)=\sqrt{P_{L}(z,k)/P_{\zeta}(k)}$ is the matter transfer function at a given redshift $z$.

Apart from the non-Gaussianities generated due to purely gravitational terms, we will also need to account for the non-linear response of the PNG to the late time power spectrum due to gravity. We follow the same EFT of LSS model described in Ref. \cite{Cabass:2022wjy}. The perturbation theory model we use in this work is consistent with the ones previously used in Refs. \cite{Nishimichi:2020tvu, Ivanov:2021kcd}. We include the effects of redshift-space distortion (RSD), Fingers-of-God (FoG) \cite{Perko:2016puo,Senatore:2014vja}, infrared (IR) resummation \cite{Senatore:2014via, Baldauf:2015xfa,Ivanov:2018gjr}, along with the appropriate counterterms at the corresponding order of perturbation theory. We rely on the publicly available code \texttt{CLASS-PT}\footnote{\url{https://github.com/Michalychforever/CLASS-PT}}\cite{Chudaykin:2020aoj}, an extension of the {\tt CLASS}\footnote{\url{https://lesgourg.github.io/class_public/class.html}}~\cite{Blas:2011rf} Boltzmann code, in order to obtain the galaxy power spectrum. 

Under the EFT of LSS framework, the perturbative expansion of galaxy density fluctuations in terms of the linear matter fluctuations $\delta_L(k,z)$ takes the following form:
\begin{eqnarray}
    \delta_g(\kvec,z) &=& Z_1(\kvec)\delta_L(k,z) + \int_{\qvec} Z_2(\qvec,\kvec-\qvec)\delta_L(\qvec,z)\delta_L(\kvec-\qvec,z) \nonumber \\
    &&+ \int_{\qvec_1,\qvec_2,\qvec_3}(2\pi)^3\delta_D(\qvec_{123}-\kvec) Z_3(\qvec_1,\qvec_2, \qvec_3)\delta_L(\qvec_1,z)\delta_L(\qvec_2,z)\delta_L(\qvec_3,z) \nonumber \\
    &&+ \fnl b_{\zeta}\frac{k^2}{\knl^2}\frac{\delta_L(\kvec,z)}{T(k,z)}+\delta_{\rm ctr}(\kvec,z),
\end{eqnarray}
Here, $k_{\rm NL}$ is the non-linear scale for matter fluctuations, which is redshift-dependent in general. To avoid clutter, we will be omitting the redshift $z$ for the transfer function and the density field from now on. The RSD kernels are given by:
\begin{eqnarray}
    Z_1(\kvec) &=& b_1 + f\mu^2\\
    Z_2(\kvec_1, \kvec_2) &=& b_1F_2(\kvec_1, \kvec_2)+\frac{b_2}{2}+b_{\mathcal{G}_2}S^2(\kvec_1, \kvec_2) + f\mu^2 G_2(\kvec_1, \kvec_2)\nonumber \\
    &&+\frac{f\mu k}{2}\left[\frac{\mu_1}{k_1}(b_1+f\mu_2^2)+\frac{\mu_2}{k_2}(b_1+f\mu_1^2)\right]\nonumber\\
S^2(\kvec_1, \kvec_2) &=& (\hat{\boldsymbol{k}}_1\cdot \hat{\boldsymbol{k}}_2)^2 - 1,
\end{eqnarray}
and $F_2(\kvec,\kvec')$ and $G_2(\kvec,\kvec')$ are the usual standard perturbation theory (SPT) kernels \cite{Bernardeau:2001qr}. The expression for $Z_3$ can be found in Ref.~\cite{Chudaykin:2020aoj}. Here, $b_1$, $b_2$ and $b_{\mathcal{G}_2}$ are the galaxy bias parameters and $f$ is the growth rate. The piece $\delta_{\rm ctr}$ contains all the required counterterm operators, which we will specify later. Lastly, the $\fnl b_{\zeta}$ term is the so-called scale-dependent bias term \cite{Dalal:2007cu,Assassi:2015fma}.

Using the ingredients above, we can write the galaxy power spectrum as: 
\begin{align}
    P_{gg}(k,\mu) &= P_{gg,L}(k,\mu) + P_{gg,{\rm PNG-bias}}(k,\mu)+ P_{gg,1L}(k,\mu)\nonumber\\
    &\quad+ P_{gg,{\rm ctr}}(k,\mu)+P_{\rm stoch}(k,\mu).
\end{align}
The linear piece $P_{gg,L}$ is the linear RSD power spectrum, which includes the damping of the baryon acoustic oscillation (BAO) feature and captures the IR resummation \cite{Ivanov:2018gjr}. The term $P_{gg,1L}$ contains the standard perturbation theory (SPT) 1-loop piece, as well as the non-Gaussian loop integral due to PNG \cite{Cabass:2022wjy}:
\begin{eqnarray}
    P_{gg,\rm{1L}}(k,\mu) &=& P_{gg,{\rm 1L\,SPT}}(k,\mu) + 2Z_1(\kvec)\int_{\qvec} Z_2(\qvec,\kvec-\qvec)B_L(\qvec,\kvec-\qvec,-\kvec).
\end{eqnarray}
Details about the 1-loop SPT term can be found in Ref. \cite{Chudaykin:2020aoj}. The term $P_{gg,{\rm PNG-bias}}$ corresponds to the well known scale-dependent bias contribution:
\begin{eqnarray}
    P_{gg,{\rm PNG-bias}}(k,\mu)&=&\frac{2\fnl b_{\zeta}Z_{1}(\k)k^2}{k_{\rm NL}^2}\frac{P_{L}(k)}{T(k)}.
\end{eqnarray}
Next, we have the counterterm contribution, $P_{gg,{\rm ctr}}$, that captures short-distance effects, and it is given by \cite{Senatore:2014vja}:
\begin{eqnarray}\label{eq:counterterms}
    P_{gg,{\rm ctr}}(k,\mu) &=& \left[\tilde{c}_0 + \tilde{c}_2 f\mu^2 + \tilde{c}_4 f^2\mu^4\right]k^2 P_L(k) - \tilde{c}f^4 \mu^4 k^4 (b_1+f\mu^2)^2 P_L(k),
\end{eqnarray}
where $\{\tilde{c}_0,\tilde{c}_2,\tilde{c}_4,\tilde{c}\}$ are free parameters to be fit from the data. Finally, there is a stochastic shot-noise contribution to the power spectrum given by:
\begin{align}
    P_{\rm stoch}(k,\mu) &= \frac{1}{\bar{n}}\left[1+P_{\rm shot}+a_0\left(\frac{k}{k_{\rm NL}}\right)^2+a_2\mu^2\left(\frac{k}{k_{\rm NL}}\right)^2\right].
\end{align}
Here, $\{P_{\rm shot}, a_0, a_2\}$ are free parameters as well, and should be included in order to account for non-Poissonian features such as exclusion effects and fiber collisions in the shot-noise. With all these pieces, the power spectrum multipoles are defined as:
\begin{eqnarray}
    P_{gg,\ell}(k) &=& \frac{2\ell+1}{2}\int_{-1}^{1}\d\mu\,\Lagr_\ell(\mu)P_{gg}(k,\mu),
\end{eqnarray}
where $\Lagr_\ell(\mu)$ is the Legendre polynomial of order $\ell$.

The bispectrum, in contrast, is more straightforward since we work at linear order in SPT:
\begin{eqnarray}\label{E:galaxy bispectrum}
    B_g(\kvec_1,\kvec_2, \kvec_3) &=& B_{g, G}(\kvec_1,\kvec_2, \kvec_3) +  Z_1(\kvec_1)Z_1(\kvec_2) Z_1(\kvec_3)B_L(\kvec_1,\kvec_2, \kvec_3)\\
    B_{g, G}(\kvec_1,\kvec_2, \kvec_3) &=& 2Z_2(\kvec_1, \kvec_2)Z_1(\kvec_1)Z_1(\kvec_2)P_L(k_1)P_L(k_2) + \text{2 perms.},
\end{eqnarray}
where $B_{g, G}$ is the galaxy bispectrum from the Gaussian initial condition. Although our bispectrum is only at tree-level, following Ref. \cite{Ivanov:2021kcd}, we choose to include separate counterterms in order to capture the FoG effects \cite{Senatore:2014vja,Perko:2016puo,Ivanov:2021kcd}. This can be taken care of by modifying the $Z_1$ kernel in Eq.~\eqref{E:galaxy bispectrum} to
\begin{eqnarray}
    Z_1(\kvec) &\to& Z_1^{\rm FoG}(\kvec) \nonumber \\
    &=&(b_1+f\mu^2 - c_0^Bk^2 -c_1^B k^2\mu^2 - c_2^B k^2 \mu^4),
\end{eqnarray}
where $\{c_0^B,c_1^B,c_2^B\}$ are additional free parameters\footnote{Note that we use different counterterms in the bispectrum, although they correspond to the same operators as those that appear in $P_{gg}(k,\mu)$. This is because the counterterms in the power spectrum renormalize the loop-correction in addition to capturing the FoG effect. Since we work at linear order only in the bispectrum, only the FoG effects need to be captured. We thank O. Philcox for clarification regarding this point.}. The shot-noise contribution to the bispectrum can be written as:
\begin{eqnarray}
    B_{\rm stoch}(\kvec_1, \kvec_2, \kvec_3) &=& \frac{B_{\rm shot}}{\bar{n}}\left[Z_1(\k_1)b_1P_{L}(k_1)+\text{2 perms.}\right]\nonumber\\
    &&+\frac{1+P_{\rm shot}}{\bar{n}}\left[2Z_1(\k_1)f\mu_1^2P_{L}(k_1)+\text{2 perms.}\right]+\frac{1+A_{\rm shot}}{\bar{n}^2},
\end{eqnarray}
where $A_{\rm shot}$ and $B_{\rm shot}$ are additional free parameters that measure the deviation from the Poisson prediction, which we must fit from the data. Lastly, to include the BAO damping, we approximate it by making the substitution $P_L(k)\to P_{L,{\rm IR}}(k)$ in all of the expressions above, i.e. the damping is included at the level of the linear matter power spectrum~\cite{DAmico:2022gki}. We provide details about this procedure in Appendix~\ref{A:IR}.

\subsection{Alcock-Paczynski Effect}

The coordinate transformation from redshift space to physical distance relies on the assumption of a fiducial cosmology. For us, that corresponds to the following choice of fiducial parameters: $\Omega_m=0.31$ and $h=0.676$ \cite{BOSS:2016hvq}. As a result, we need to perform an Alcock-Paczynski correction~\cite{Alcock:1979mp} in order to make the coordinate system consistent between the observation and the theory. Following the analysis in Ref. \cite{DAmico:2022gki}, we account for the Alcock-Pacynski effect only at the tree-level, as it is estimated to be negligible at the one-loop level.

First, let us define the dilation parameters~\cite{Gil-Marin:2016wya,Chudaykin:2020aoj}:
\begin{align}
    \alpha_{||}(z) &= \frac{H^{\rm fid}(z)}{H(z)},\nonumber\\
    \alpha_{\perp}(z) &= \frac{D_{A}(z)}{D_{A}^{\rm fid}(z)}\,,
\end{align}
where $D_A(z)$ is the angular diameter distance to redshift $z$. If we set $F = \alpha_{||}/\alpha_{\perp}$, then we can write the observed bispectrum and power spectrum in terms of the ones computed using the fiducial cosmology as:
\begin{align}
    B^{\rm obs}_g(\tilde{k}_1,\tilde{k}_2,\tilde{k}_3,\tilde{\mu}_1,\tilde{\mu}_2) &= \frac{1}{\alpha_{||}^2\alpha_{\perp}^4}B_g(k_1,k_2,k_3,\mu_1,\mu_2)\,,\nonumber\\
    P^{\rm obs}_{gg}(\tilde{k},\tilde{\mu}) &= \frac{1}{\alpha_{||}\alpha_{\perp}^2}P_{gg}(k,\mu)\,,
\end{align}
where $k$ and $\mu$ refer to the fiducial cosmology, and $\tilde{k}$ and $\tilde{\mu}$ refer to the observed cosmology, and they are related through the following relation:
\begin{align}
    k &= \frac{\tilde{k}}{\alpha_{\perp}}\left[1+\tilde{\mu}^2(F^{-2}-1)\right]^{1/2}\,,\nonumber\\
    \mu &= \frac{\tilde{\mu}}{F}\left[1+\tilde{\mu}^2(F^{-2}-1)\right]^{-1/2}\,.
\end{align}

\subsection{Survey Geometry}\label{sec:surveyGeometry}

Due to the non-trivial survey geometry, the theory power spectrum and bispectrum must be appropriately convolved with the survey window function in order to match the one measured from the data. Here we outline the procedure that we follow to account for the survey geometry for the power spectrum multipoles and the bispectrum shapes. 

The window convolution for the power spectrum can be streamlined by working with the two-point correlation function, computed using Hankel transforms~\cite{BOSS:2016hvq,BOSS:2016psr}. That is, if we have the unconvolved power spectrum multipoles $P_{\ell}^{\rm unconv}(k)$, we can first compute the Hankel transform\footnote{The Hankel transforms are performed with \texttt{hankl}\cite{Karamanis:2021xbv} using the FFTLog algorithm~\cite{Hamilton:1999uv}. This can be found at \url{https://hankl.readthedocs.io/en/latest/}.} to obtain the two-point correlation function multipoles $\xi_\ell(r)$, and then convolve them with the window function multipoles $W_\ell(r)$\footnote{The $W_\ell(r)$s are publicly available at \url{https://www.ub.edu/bispectrum/page11.html} for the BOSS/{\tt Patchy} [CMASS, LOWZ]$\times$[NGC, SGC] simulations, as well as the ones for {\tt Nseries} simulations.} to get $\hat{\xi}_\ell(r)$:
\begin{align}
    \hat{\xi}_0(r) &= \xi_{0}(r)W_0(r)^2 + \frac{1}{5}\xi_{2}(r)W_2(r)^2 + \frac{1}{9}\xi_{4}(r)W_4(r)^2 \\
    \hat{\xi}_2(r) &= \xi_{0}(r)W_2(r)^2 + \xi_{2}(r)\left[W_0(r)^2+\frac{2}{7}W_2(r)^2+\frac{2}{7}W_4(r)^2\right]\nonumber\\
    &+ \xi_{4}(r)\left[\frac{2}{7}W_2(r)^2+\frac{100}{693}W_4(r)^2\right]\\
    \hat{\xi}_4(r) &= \xi_0(r)W_4(r)^2 + \xi_2(r)\left[\frac{18}{35}W_2(r)^2+\frac{20}{77}W_4(r)^2\right]\nonumber\\
    &+ \xi_4(r)\left[W_0(r)^2+\frac{20}{77}W_2(r)^2+\frac{162}{1001}W_4(r)^2\right].
\end{align}
Finally, the window-convolved power spectrum multipoles $P_\ell^{\rm conv}(k)$ can be obtained by performing the inverse Hankel transform with $\hat{\xi}_\ell(r)$:
\begin{align}
    P_{\ell}^{\rm conv}(k) &= 4\pi(-i)^{\ell}\int\d r\,r^2\hat{\xi}_{\ell}(r)j_{\ell}(kr)\,.
\end{align}
Due to convergence issues with the Hankel transform, this procedure must be modified slightly if $P_\ell(k)$ is at one-loop order instead, or does not decay sufficiently fast at large $k$. We artificially extrapolate the power spectra at high $k$ ($k\geq0.5$ $h$Mpc$^{-1}$) to be proportional to $k^{-4}$, so that the Hankel transform converges.

In the bispectrum case, directly convolving with the window function is more computationally expensive. To overcome this issue, we use an approximate method following Refs.~\cite{Gil-Marin:2014sta,Gil-Marin:2016wya}. Suppose the functional form of the galaxy bispectrum is a product of two matter power spectra and a function of wavenumbers $\k$,
\begin{align}
    B_{g}(\k_1,\k_2,\k_3) &= P_L(k_1)P_L(k_2)\mathcal{Q}(\k_1,\k_2,\k_3)+\text{2 perms.}\,,
\end{align}
then the window-convolved bispectrum can be approximated as:
\begin{align}
    &\int_{\k',\k''}B_g(\k',\k'',-\k'-\k'')W_{3}(\k_1-\k',\k_2-\k'') \nonumber\\
    &\qquad\qquad\qquad\approx \left[P_L\star W\right](\k_1) \left[P_L\star W\right](\k_2) \mathcal{Q}(\k_1,\k_2,\k_3)+\text{2 perms.}\,,
\end{align}
where $W_3(\kvec_1,\kvec_2)\equiv \frac{1}{I_{33}}m(\kvec_1)m(\kvec_2)m^*(\kvec_1+\kvec_2)$, $W(\k)=\frac{1}{I_{22}}|m(\k)|^2$, $m(\kvec)$ is the Fourier transform of the mask, and the normalization factors $I_{22}$ and $I_{33}$ will be precisely defined in \S\ref{sec:galaxyCataogues} \cite{Gil-Marin:2014sta}. The notation $P_L\star W$ denotes the window convolution applied to the power spectrum, as discussed previously in this section. This approximation is valid for modes that are not comparable to the survey geometry. For wavenumbers $\gtrsim 0.03\,h$Mpc$^{-1}$, the error of the approximation is less than 5\%, with only 1\% differences on parameter estimation using modes between 0.03 to 0.15 $h$Mpc$^{-1}$ \cite{Gil-Marin:2014sta}. Following Ref. \cite{DAmico:2022gki} we employ a similar approximation to convolve the primordial piece:
\begin{align}
    B_{L}(\k_1,\k_2,\k_3) &= \frac{1}{3}\left(\frac{B_{L}(\k_1,\k_2,\k_3)}{P_L(k_1)P_L(k_2)}P_L(k_1)P_L(k_2)+2\text{ perms.}\right)\nonumber\\
    &\rightarrow \frac{1}{3}\left(\frac{B_{L}(\k_1,\k_2,\k_3)}{P_L(k_1)P_L(k_2)}(W\star P_L)(\k_1)(W\star P_L)(\k_2)+2\text{ perms.}\right).
\end{align}

We find that the error due to this approximation, across all skew-spectra, is at most $6\%$ for the equilateral shape and $18\%$ for the orthogonal shape. Details regarding these estimates can be found in Appendix~\ref{A:validation}. We will further assess the goodness of this approximation through a validation test of our pipeline on the {\tt Nseries} simulations in \S\ref{sec: results}. 

\subsection{Skew-Spectrum}\label{sec:skewspec_kernels}

In this section, we will describe the formalism of the skew-spectra used in our analysis of the BOSS data. In order to obtain the theory skew-spectrum, we must specify the kernels used, $\mathcal{S}(\kvec,\kvec')$. From our previous work \cite{Chakraborty:2022aok}, we know that the Legendre polynomial kernels can preserve much of the information from the full bispectrum analysis and, therefore, these will be our baseline kernels. In addition, we will add the primordial non-Gaussianity shapes for both equilateral and orthogonal as extra kernels. To sum up, our skew-spectra are defined as follows:
\begin{align}\label{E:skewspectrum}
    \tilde{P}_{\mathcal{S}_i}(k) = \int\frac{\d\hat{\kvec}}{4\pi} \int_{\q} \mathcal{S}_{i}(\q,\k-\q)B_{g}(\q,\k-\q,-\k),
\end{align}
with kernels $\mathcal{S}_i$ given by:
\begin{align}
    \mathcal{S}_1 &= 1\\
    \mathcal{S}_2 &= \frac{1}{2}\left(\frac{k_1}{k_2}+\frac{k_2}{k_1}\right)\frac{\k_1\cdot\k_2}{k_1k_2}\\
    \mathcal{S}_3 &= \frac{1}{2}\left[3\left(\frac{\k_1\cdot\k_2}{k_1k_2}\right)^2-1\right]\\
    \mathcal{S}_{\rm equil} &= T(k_1)T(k_2)T(k_3)\Delta_{\zeta}^4\frac{\mathcal{B}_{\rm equil}(k_1,k_2,k_3)}{k_1^2k_2^2k_3^2}\times\frac{1}{P_L(k_1)P_L(k_2)}\\
    \mathcal{S}_{\rm ortho} &= T(k_1)T(k_2)T(k_3)\Delta_{\zeta}^4\frac{\mathcal{B}_{\rm ortho}(k_1,k_2,k_3)}{k_1^2k_2^2k_3^2}\times\frac{1}{P_L(k_1)P_L(k_2)}.
\end{align}

In order to perform the integral in Eq.~\eqref{E:skewspectrum}, we choose the following coordinate system \cite{Schmittfull:2020hoi}:
\begin{equation}
    \hat{\bm{z}} = (0,0,1),\qquad \hat{\bm{k}}=\left(0, \sqrt{1-\mu_k^2},\mu_k\right),\qquad \hat{\bm{q}}=\left(\sqrt{1-\mu_q^2}\cos\phi_q,\sqrt{1-\mu_q^2}\sin\phi_q,\mu_q\right),
\end{equation}
where $\mu_k=\hat{\kvec}\cdot\hat{\bm{z}}$, $\mu_q=\hat{\qvec}\cdot\hat{\bm{z}}$, $\cos(\phi_q)=\hat{\qvec}\cdot\hat{\bm{x}}$, and $\hat{\bm{z}}$ is the line-of-sight direction.
We use the publicly available package {\tt pycuba}\footnote{{\tt pycuba}, {\url{https://github.com/JohannesBuchner/PyMultiNest/}}, is a python wrapper for {\tt cuba}~\cite{Hahn:2004fe} which provides four different routines to compute the numerical integral. We use the Divonne routine for most of the computation, and have checked that it agrees with other routines for the case considered here.}~\cite{Hahn:2004fe,Buchner:2014nha} to numerically evaluate the resulting 4-dimensional integral. The skew-spectrum computed from the data is averaged over a spherical shell in Fourier space. To mimic this, the theoretical skew-spectrum $\tilde{P}_{\mathcal{S}_j}$ is computed at the $i^{\rm th}$-bin $\Delta_i$ as:
\begin{align}
    \tilde{P}_{\mathcal{S}_j}(k_{{\rm eff,}i}) = \frac{\sum_{k\in\Delta_i} k \tilde{P}_{\mathcal{S}_j}(k)}{\sum_{k\in\Delta_i}k},
\end{align}
with $k_{{\rm eff},i} = \left(\sum_{k\in\Delta_i} k^2\right)/\left(\sum_{k\in\Delta_i}k\right)$ being the effective wavenumber .

\section{The Data}\label{sec: data}

In this section, we will specify the details of our primary dataset, the $12^{th}$ data release (DR12) of the BOSS survey. We further use two simulation suites, the {\tt MultiDark-Patchy} and the {\tt Nseries} mocks for covariance estimation and validation, respectively. Finally, we will discuss the estimation of the skew-spectra from the data.

\subsection{Galaxy Catalogues}\label{sec:galaxyCataogues}

We use three different galaxy catalogues in our analysis: BOSS DR12~\cite{SDSS-III:2015hof}, {\tt MultiDark}-{\tt Patchy}\footnote{The BOSS DR12 catalogues and {\tt Patchy} simulations can be downloaded from \url{https://data.sdss.org/sas/dr12/boss/lss/}.~\cite{Kitaura:2015uqa,Rodriguez-Torres:2015vqa}. We use {\tt CMASSLOWZTOT} for BOSS data in our analysis.} (hereafter, ``{\tt Patchy} mocks") and {\tt Nseries} mocks\footnote{The {\tt Nseries} mocks can be found in \url{https://www.ub.edu/bispectrum/page11.html}.} \cite{BOSS:2016wmc,Hahn:2016kiy}. The data are broken down into two patches of the sky: the northern galactic cap (NGC) and the southern galactic cap (SGC). Further, the redshift range is cut into the redshift ranges $z_{\rm CMASS}$ and $z_{\rm LOWZ}$:
\begin{align}
    z_{\rm CMASS} &= [0.43,\,0.70]\text{ with }z_{\rm eff}=0.57\nonumber,\\
    z_{\rm LOWZ} &= [0.15,\,0.43]\text{ with }z_{\rm eff}=0.32\,.
\end{align}
The effective redshift $z_{\rm eff}$ for each range is used to paint the data onto a grid in redshift space, and also to compute the theory power spectra and skew-spectra. This turns out to be four different chunks of the sky in total for our full BOSS analysis. The survey window function is obtained via the public {\tt MANGLE} mask\footnote{Available at \url{https://space.mit.edu/~molly/mangle/}.}~\cite{Swanson:2007aj}. We give here a brief summary of the catalogues that we use: 
\begin{itemize}
    \item BOSS DR12: This is the primary catalogue that we use for parameter inference.
    \item {\tt Patchy} mocks: A set of 2048 simulations from the ``{\tt MultiDark-Patchy}" suite, which is run using an approximate scheme. We use this set to obtain the covariance matrix for our likelihood function.
    \item {\tt Nseries} mocks: We use this set of simulations to validate our analysis pipeline, as this set is a full N-body simulation unlike the approximate scheme applied in {\tt Patchy} mocks. There are a total of 84 semi-independent mocks, created from different projections of seven independent N-body simulations. Each of these $84$ boxes cover as much volume as the CMASS NGC sample.
\end{itemize}
The fiducial values of the cosmological parameters for {\tt Nseries} and {\tt Patchy} mocks are listed in Table~\ref{tab:fiducialCosmo}, together with the \textit{Planck} fiducial values, for comparison.
\begin{table}[]
    \centering
    \begin{tabular}{c|cccccc}
         & $h$ & $\omega_b$ & $\omega_c$ & $\sigma_8$ & $n_s$ & $\sum m_\nu$ [eV] \\
        \hline
        \textit{Planck} 2018 & 0.673 & 0.022 & 0.120 & 0.8120 & 0.966 & 0.06\\
        \hline
        {\tt Nseries} mocks & 0.700 & 0.023 & 0.117 & 0.8200 & 0.960 & 0.00\\
        \hline
        {\tt Patchy} mocks & 0.678 & 0.022 & 0.119 & 0.8288 & 0.960 & 0.00
    \end{tabular}
    \caption{Fiducial cosmology for \textit{Planck} 2018 data, {\tt Nseries} and {\tt Patchy} simulations.}
    \label{tab:fiducialCosmo}
\end{table}

We apply the Feldman-Kaiser-Peacock estimator (FKP)~\cite{Feldman:1993ky} to estimate power spectrum multipoles and skew-spectra in our analysis. The FKP formalism aims to minimize the statistical error by giving a weight, $\omega_{\rm FKP}=1/(1+\bar{n}_g(z)P_{\rm FKP})$, to each object in the catalogue, with $\bar{n}_g(z)$ being the number density of galaxies at redshift $z$. $P_{\rm FKP}$ is chosen to be the amplitude of the galaxy power spectrum where the signal is maximized. In our work, we set $P_{\rm FKP}=10^4\,h^{-3}\text{Mpc}^{3}$, corresponding to $k=0.10\,h$ $\text{Mpc}^{-1}$ for both BOSS data and {\tt Patchy} mocks; while for {\tt Nseries} mocks, we instead use the FKP weight provided in the public catalogue file. Additional weights $\omega_{\rm c}$ are needed in order to capture observational effects. For BOSS data, we use $\omega_{\rm c}=(\omega_{\rm rf}+\omega_{\rm fc}-1)\omega_{\rm sys}$ to capture redshift-failure, fiber collision and systematic effects~\cite{BOSS:2016psr}. For {\tt Patchy} mocks, we use $\omega_{\rm c}=\omega_{\rm veto}\omega_{\rm fc}$, for veto masks and fiber collisions. For {\tt Nseries} mocks, we use $\omega_{\rm c}$ directly provided in the catalogue file. The FKP-weighted field is then given by taking the difference between number densities of galaxies $n_g(\r)$ and that of random particles $n_r(\r)$:
\begin{align}
    F_{\lambda}(\r)=\frac{\omega_{\rm FKP}(\boldsymbol{r})}{I_{\lambda\lambda}^{1/\lambda}}\left[\omega_c(\boldsymbol{r}) n_g(\r)-\alpha n_r(\r)\right]\,,
\end{align}
where 
\begin{align}\label{eq:FKPNormalization}
    I_{\lambda_1\lambda_2}=\int\d^3\r\,\omega_{\rm FKP}^{\lambda_1}(\boldsymbol{r})\langle n_g\omega_c\rangle^{\lambda_2}(\r)\,,
\end{align}
and $\alpha$ denotes the ratio of the number between galaxies and random particles. We choose random catalogues that are 50 times larger than the galaxy dataset/mocks, in order to capture the survey geometry as well as to reduce the additional shot-noise from these random particles. The $\lambda$ in the above expression is an integer that depends on the spectra we are interested in. For example, we have $\lambda=2$ for the power spectrum and $\lambda=3$ for the bispectrum. In our case, since the skew-spectrum is derived from the bispectrum, we will adopt $\lambda=3$ for it.

To account for the survey geometry in the theory, we compute the window multipoles $W_\ell$ (see \S\ref{sec:surveyGeometry}) from random catalogues, and normalize them appropriately so that $W_0(r\rightarrow0)\rightarrow1$. Physically, this is to ensure that the small-scale power spectrum and bispectrum are unaffected by the mask. 

Particular care has to be taken in the normalization of the FKP field, i.e. the factor $I_{22}$, on the data side. The usual approximation for $I_{22}$ involves summing over all the random particles, $I_{22}^{\rm approx}\approx\alpha\sum_i \omega_{\rm FKP,i}^2n_r(\r_i)$. However, it is known that this can result in roughly $10\%$ of suppression on the amplitude of BOSS data, which leads to an incorrect estimation of $\sigma_8$~\cite{Beutler:2021eqq,deMattia:2020fkb}. In order to have consistent normalization between the data and the theoretical model, we do not use this approximation and instead perform the full random pair count with the code {\tt Corrfunc} \cite{Sinha:2019reo}\footnote{\url{https://corrfunc.readthedocs.io/en/master/}}. The correct $I_{22}$ can be evaluated with the random pair count (pc) as (see e.g. \cite{Philcox:2019hdi}):
\begin{align}
    I_{22}^{\text{pc}}=\frac{3\alpha^2 RR(\Delta)}{4\pi(r_{\rm max}^3-r_{\rm min}^3)}\,,
\end{align}
where $\Delta$ is the $r$-bin centered at $1\,h^{-1}\text{Mpc}$ with range being $[r_{\rm min},\,r_{\rm max}]$ and $RR(r)$ is the random pair count including FKP weights with separation $r$. In the end, $I_{22}^{\rm pc}$ is used to ensure consistent normalization of window multipoles between data and theoretical modelling. Our computation agrees with the one shown in Ref.~\cite{Beutler:2021eqq} for BOSS data, which split the redshift range into $z_1=[0.2,\,0.5]$ and $z_3=[0.5,\,0.75]$. 

Finally, we embed the FKP field into a cubic box with length $3500\,h^{-1}\text{Mpc}$ on each side for all the data and simulations we consider. We then apply the triangular-shaped cloud (TSC) scheme to generate a mesh with $512^3$ pixels in total, which is equivalent to having a Nyquist wavenumber of $k_{\rm Nyq}=0.46\,h\text{Mpc}^{-1}$. We also perform grid-deconvolution in Fourier space to correct aliasing from the mass assignment~\cite{Jing:2004fq}, that is, the way we assign each particle to a specific pixel. 

\subsection{Estimation of Power Spectra and Skew-Spectra}

We compute the power spectrum multipoles $\hat{P}_{\ell}(k)$ with $\ell=0,2,4$ assuming the Yamamoto approximation~\cite{Yamamoto:2005dz}, which fixes the line-of-sight (LoS) direction to that of a single galaxy for each galaxy pair, as opposed to a single LoS for the whole survey. Note that the hat here denotes that the spectrum is estimated from data or simulations. To compute the power spectrum, we choose $[k_{\rm min}^{P},\,k_{\rm max}^{P}]=[0.01,0.22]$ $h\text{Mpc}^{-1}$ with bin size $\Delta k^{P}=0.0115$ $h\text{Mpc}^{-1}$ for both the data and the simulations.

The separability of the skew-spectra kernels enables us to expedite their estimation from the data. In particular, given a kernel that can be decomposed as $\mathcal{S}(\q_1,\q_2)=\sum_i f_i(\q_1)g_i(\q_2)h_i(-\k)$ with $\q_1+\q_2=\k$, the skew-spectrum can be written as a sum:
\begin{align}\label{eq:skewSpectrumEst}
    \hat{\tilde{P}}_{\mathcal{S}}(k) = \sum_i\int\frac{\d\hat{\k}}{4\pi}\left[f_i F_3 * g_i F_3\right](\k)\left[h_i F_3\right](-\k)\,,
\end{align}
where $F_3(\r)$ is the FKP field defined in \S\ref{sec:galaxyCataogues} with $\lambda=3$. The convolution, being the most expensive part of the above expression, can easily be recasted into a sequence of FFTs by relying on the fact that convolution in Fourier space is a point-wise product in position space. Therefore, after we have the FKP density field on the grid, we can then compute the skew-spectra by following these steps:
\begin{itemize}
    \item Transform the FKP field into Fourier space: $F_3(\r)\rightarrow F_3(\k)$.
    \item Compute $f_i(\k)F_3(\k)$, $g_i(\k)F_3(\k)$ and $h_i(\k)F_3(\k)$ by point-wise products in Fourier space.
    \item Transform the data back to position space, and get $\left[f_i F_3\right](\r)$, $\left[g_i F_3\right](\r)$ and $\left[h_i F_3\right](\r)$. 
    \item Compute point-wise products between $\left[f_i F_3\right](\r)$ and $\left[g_i F_3\right](\r)$. This is equivalent to doing convolution in Fourier space and we denote it as $\left[f_i F_3\star g_i F_3\right](\r)$.
    \item Compute the cross-correlation between $\left[f_i F_3 \star g_i F_3\right](\r)$ and $\left[h_i F_3\right](\r)$ and sum over $i$, which runs over the components of the skew-spectrum, as shown in Eq.~\eqref{eq:skewSpectrumEst}. We can then obtain the targeted skew-spectrum.
\end{itemize}

All the kernels for the skew-spectrum that we consider here are separable (see \S\ref{sec:skewspec_kernels}), and hence we can follow the above steps to expedite their computation. In order to estimate the skew-spectra from the data, we must specify two different ranges of $k$. The first range of $k$ is the range in which we perform the convolution. We set this range to be $\mathcal{I}_{\rm conv}=[k_{\rm conv,min}^{\tilde{P}},\,k_{\rm conv,max}^{\tilde{P}}]=[0.03,\,0.08]$ $h\text{Mpc}^{-1}$, in which the density field lies within the linear regime. The choice of $k_{\rm conv,min}^{\tilde{P}}$ also ensures that our approximation for the window convolution of the bispectrum (mentioned in \S\ref{sec:surveyGeometry}) is within $5\%$ error. The second range of $k$ is the range over which we evaluate the cross-correlation between $\left[f_i F_3* g_i F_3\right](\r)$ and $\left[h_i F_3\right](\r)$. We set this range to be $\mathcal{I}_{\rm skew}=[k_{\rm min}^{\tilde{P}},\,k_{\rm max}^{\tilde{P}}]=[0.035,\,0.075]$ $h\text{Mpc}^{-1}$ with bin-width $\Delta k^{\tilde{P}}$ chosen to be 0.01 $h\text{Mpc}^{-1}$. Note that these choices ensure $\mathcal{I}_{\rm skew}\subset\mathcal{I}_{\rm conv}$ in order to discard modes on the boundary of $\mathcal{I}_{\rm conv}$, which suffer larger numerical errors from the convolution.

For both power spectrum multipoles and skew-spectra, we subtract the expected Poisson shot-noise terms in order to minimize the potentially different noise features between the data and the simulations. The shot-noise of the power spectrum multipoles is given by: $\hat{P}^{\rm shot}(k)=(1+\alpha)I_{21}/I_{22}$. For the skew-spectra, we adopt the shot-noise contribution for the bispectrum~\cite{Gil-Marin:2014sta}, and follow Ref.~\cite{Hou:2022rcd} to adapt it for the skew-spectra:
\begin{align}
    \hat{\tilde{P}}^{\rm shot}_{\mathcal{S}}(k)&=\sum_i \int\frac{\d\hat{\k}}{4\pi} h_i(-\k)\left [\left(\frac{I_{31}}{I_{33}}+\frac{I_{32}\hat{P}(\k)}{I_{33}}\right)[f_i\star g_i](\k)+\frac{2I_{32}}{I_{33}}[\hat{P}f_i \star g_i](\k)\right]\,.
\end{align}
where $\hat{P}(\k)$ is the power spectrum estimated from data. Note that $I_{\lambda_1\lambda_2}$ here is the same as the one in Eq.~\eqref{eq:FKPNormalization}. Despite these subtractions, we maintain nuisance parameters in our theory model, which measure the deviation from the Poisson prediction, and so the priors for nuisance parameters of the shot-noise will therefore be centered at zero shot-noise.

We use the code {\tt nbodykit}\footnote{\url{https://nbodykit.readthedocs.io/en/latest/}.}~\cite{Hand:2017pqn} to paint the catalogues onto a grid, as well as to facilitate our estimation of these spectra. In particular, we use the functions {\tt ConvolvedFFTPower()} to estimate the power spectrum multipoles and {\tt FFTPower()} to compute the skew-spectra (specifically, in the cross-correlation between the quadratic field and the linear field).

\section{Parameter Inference}\label{sec: inference}

In this section, we will discuss details of our parameter inference and describe the choices we have made. We start with a discussion of our likelihood function and the covariance matrix, followed by a discussion of our sampling method and prior choices. 

\subsection{Likelihood Function}\label{sec:covarianceMatrix}

Working with the assumption of a Gaussian likelihood, we compute the covariance $\hat{\mathsf{C}}$ from 2048 {\tt Patchy} mocks. 

Our complete data vector is $\boldsymbol{d}^i=\left\{P_0^i,P_2^i,P_4^i,\skew_0^i,\skew_1^i,\skew_2^i,\skew_{\rm equil}^i,\skew_{\rm ortho}^i\right\}$, i.e. the power spectrum multipoles and skew-spectra with multiple shapes (where $i$ here denotes the index for the simulation). In order to compute the covariance matrix, we first compute the quantity $\hat{\mathsf{C}}^{*}$:
\begin{align}
    \hat{\mathsf{C}}^{*} = \frac{1}{n-1}\sum_{i=1}^{n}(\boldsymbol{d}^i-\bar{\boldsymbol{d}})(\boldsymbol{d}^i-\bar{\boldsymbol{d}})^{\top},
\end{align}
where $n=2048$ is the number of simulations and $\bar{\boldsymbol{d}}$ is the sample mean of the data vector. We further include the Hartlap factor to make it an unbiased estimate of the covariance~\cite{Hartlap:2006kj},
\begin{align}
    \hat{\mathsf{C}} = \frac{n-1}{n-p-2}\hat{\mathsf{C}}^*\,,
\end{align}
where $p$ denotes the number of data points, or simply the length of the data vector $\boldsymbol{d}^i$.

As we assume a Gaussian likelihood, we perform the matrix inversion for the covariance matrix as follows. First, we separate the covariance matrix into a diagonal and an off-diagonal part: $\hat{\mathsf{C}}=\hat{\mathsf{C}}^{\rm diag}+\hat{\mathsf{C}}^{\rm offdiag}$. The piece $\hat{\mathsf{C}}^{\rm diag}$ is block diagonal, i.e. it contains the diagonal part of the covariance between all pairs of $\hat{P}_\ell$ and $\hat{\tildeP}_{\mathcal{S}_i}$, while the off-diagonal part $\hat{\mathsf{C}}^{\rm offdiag}$ contains all of the cross-correlation across different $k$-bins. 

Next, we apply Singular Value Decomposition (SVD)\footnote{SVD is a general decomposition wherein a matrix $M$ may be written as a product $U S V$, in terms of unitary matrices $U$ and $V$ and a rectangular diagonal matrix $S$.} to the non-diagonal part: $\hat{\mathsf{C}}^{\rm offdiag}=USV$, where $S$ is a square diagonal matrix and both $U$ and $V$ are unitary matrices. Using the Woodbury formula\footnote{$(A+UCV)^{-1} = A^{-1}-A^{-1}UC(I+VA^{-1}UC)^{-1}VA^{-1}$}, the inverse of the covariance matrix can be written down as~\cite{Lacasa:2022fxk}:
\begin{align}
    \hat{\mathsf{C}}^{-1} = (\hat{\mathsf{C}}^{\rm diag})^{-1} - (\hat{\mathsf{C}}^{\rm diag})^{-1}U S\left(I+V(\hat{\mathsf{C}}^{\rm diag})^{-1}U S\right)^{-1}V(\hat{\mathsf{C}}^{\rm diag})^{-1}\,.
\end{align}
As a result, the inverse of the covariance now turns out to be a correction to the inverse of the diagonal matrix. Additionally, we note that the matrix $\hat{\mathsf{C}}^{\rm offdiag}$ is of low rank, and the Woodbury formula enables us to perform the low-rank approximation of the inverse\footnote{The idea of low-rank approximation is that we can get rid of small eigenvalues in $S$ inside the SVD decomposition. That is, we set $S_{ii}=0$ if $\max_{i'} S_{i'i'}/S_{ii}>c_{\rm LR}$ for some threshold $c_{\rm LR}$ of our choice.}. This procedure can help us reduce small numerical fluctuations in the inverse as well as speed-up the matrix product in the likelihood function. 

With the Woodbury simplified covariance, the negative log-likelihood function can be broken down into diagonal and off-diagonal pieces:
\begin{align}
    -2\log\mathcal{L} &= -2\log\mathcal{L}_{\rm diag}-2\log\mathcal{L}_{\rm offdiag}.
\end{align}
Here, $\mathcal{L}_{\rm diag}$ and $\mathcal{L}_{\rm offdiag}$ correspond to:
\begin{align}
    -2\log\mathcal{L}_{\rm diag} &= (\boldsymbol{d}-\boldsymbol{m})^\top(\hat{\mathsf{C}}^{\rm diag})^{-1}(\boldsymbol{d}-\boldsymbol{m}),\nonumber\\
    -2\log\mathcal{L}_{\rm offdiag} &= -\left[U^\top(\hat{\mathsf{C}}^{\rm diag})^{-1}(\boldsymbol{d}-\boldsymbol{m})\right]^\top S \left(I+V(\hat{\mathsf{C}}^{\rm diag})^{-1}U S\right)^{-1}\left[V(\hat{\mathsf{C}}^{\rm diag})^{-1}(\boldsymbol{d}-\boldsymbol{m})\right]\,,
\end{align}
where $\boldsymbol{d}$ is the data vector being either BOSS data or {\tt Nseries} mocks, and $\boldsymbol{m}$ is the vector containing the theoretically computed power spectrum multipoles and skew-spectra.

\subsection{Parameters and Prior Choices}

In our analysis, we assume a $\Lambda$CDM cosmology with parameters fixed to \textit{Planck} 2018 best-fit values~\cite{Planck:2018vyg} and we set the sum of neutrino mass to be 60 meV. As a result, in addition to $\fnle$ and $\fnlo$, we vary the EFT parameters, i.e. the galaxy bias and the counterterms, as well as nuisance parameters for shot-noise in our analysis. We follow Refs. \cite{Philcox:2021kcw,Ivanov:2023qzb,Cabass:2022wjy} for the choices of priors, which are motivated by previous BOSS analyses for the counterterms\footnote{The counterterms at one loop ($c_{0,2,4}$ and $c_{0,1,2}^{\rm B}$) are expected to be $\mathcal{O}(1)\times \knl^{-2}$. However, due to the strong FoG effect already detected in the BOSS sample we expect it to take a larger value than the naive EFT estimate, and the priors are taken to be wide enough to accommodate for this. Meanwhile, although $\tilde{c}$ (in Eq. \eqref{eq:counterterms}) is at two-loop, previous power spectrum analyses detect a value much larger ($\sim 500$ [Mpc/h]${}^4$) than the EFT estimate, and the prior is taken accordingly \cite{Chudaykin:2020aoj}.} and by the co-evolution model \cite{Desjacques:2016bnm} for the bias parameters. The priors for bias parameters are set up to be:
\begin{align}
    &b_1\sim \text{Uniform}(0,4)\,,\quad b_2\sim\text{Uniform}(-4,4)\,,\quad b_{\mathcal{G}_2}\sim\text{Uniform}(-4,4)\,,\nonumber\\
    &b_{\Gamma_3}\sim\mathcal{N}(\frac{23}{42}(b_1-1),1^2)\,,
\end{align}
where $\mathcal{N}(\mu,\sigma^2)$ is a normal distribution with mean $\mu$ and variance $\sigma^2$, and ${\rm Uniform}(a,b)$ sets a flat prior between the bounds $a$ and $b$.

For the counterterms and the shot-noise parameters, we have:
\begin{align}
    &\frac{c_0}{[\text{Mpc}/h]^2}\sim\mathcal{N}(0,30^2)\,,\quad \frac{c_2}{[\text{Mpc}/h]^2}\sim\mathcal{N}(30,30^2)\,, \quad
    \frac{c_4}{[\text{Mpc}/h]^2}\sim\mathcal{N}(0,30^2)\,, \nonumber\\
    &\frac{\tilde{c}}{[\text{Mpc}/h]^4}\sim\mathcal{N}(500,500^2)\,, \quad \frac{c_1^{\rm B}}{[\text{Mpc}/h]^2}\sim\mathcal{N}(0,50^2)\,, \quad \frac{c_0^{\rm B}}{[\text{Mpc}/h]^2}\sim\mathcal{N}(0,50^2)\,, \nonumber\\
    & \frac{c_2^{\rm B}}{[\text{Mpc}/h]^2}\sim\mathcal{N}(0,50^2)\,, \quad A_{\rm shot}\sim\mathcal{N}(-1,1^2)\,, \quad P_{\rm shot}\sim\mathcal{N}(-1,2^2)\,, \nonumber\\
    &\quad B_{\rm shot}\sim\mathcal{N}(0,1^2)\,,\quad a_0\sim\mathcal{N}(0,2^2)\,,\quad a_2\sim\mathcal{N}(0,2^2)\,.
\end{align}
The counterterms $\left\{c_0,\,c_2,\,c_4\right\}$ here are related to $\left\{\tilde{c}_0,\,\tilde{c}_2,\,\tilde{c}_4\right\}$ (defined in \S\ref{sec: theory}) by the mapping~\cite{Chudaykin:2020aoj}: $c_0\equiv\tilde{c}_0+f\tilde{c}_2/3+f^2\tilde{c}_4/5$, $c_2\equiv\tfrac{3}{4}(\tilde{c}_2+6f\tilde{c}_4/7)$ and $c_4\equiv\tfrac{8}{35}\tilde{c}_4$\footnote{This mapping accounts for the change of basis of the counterterm parameters as one computes the multipoles of $P_{gg,{\rm ctr}}(k,\mu)$ (see Eq (\ref{eq:counterterms})). In other words, it specifies the combination of the $\tilde{c}_\ell$ coefficients appearing in the power spectrum multipoles $P_{gg,\ell}(k)$. For example, $P_{gg,0}^{\rm ctr}(k)=(\tilde{c}_0+f\tilde{c}_2/3+f^2\tilde{c}_4/5)k^2 P_L(k)\equiv c_0 k^2 P_L(k)$.}. We also note here that we include two additional parameters $c_0^{B}$ and $c_2^{B}$ for the skew-spectra (compared with a single counterterm as in  Refs.~\cite{Philcox:2021kcw,Cabass:2022wjy}), since our skew-spectra are sensitive to RSD information. For the primordial non-Gaussianity parameters, $\fnle$ and $\fnlo$, we adopt uninformative flat priors. In practice, we have
\begin{align}
    \fnl^{\rm equil} \sim \text{Uniform}(-2000,2000)\,, \quad \fnl^{\rm ortho} \sim \text{Uniform}(-1000,1000)\,.
\end{align}
For the non-local scale-dependent bias $b_\zeta$, we have \cite{Cabass:2022wjy}:
\begin{align}
    b_\zeta = \bar{b}_\zeta \frac{18}{5}\delta_c(b_1-1)\text{  with  }\bar{b}_\zeta\sim\mathcal{N}(1,5)\text{ and }\delta_c=1.686\,.
\end{align}
To sum up, we have a total of 17 nuisance parameters in each patch of the sky, so the total number of parameters are $17\times N_{\rm patch}+2$, with $N_{\rm patch}$ being the number of patches. Since we use both CMASS and LOWZ redshift cuts, each with NGC and SGC skies, we have $N_{\rm patch}=4$, giving us a total of $70$ parameters. 

Finally, in order to sample our posterior, we will use nested sampling~\cite{Skilling:2006gxv}, which will be done with the public package {\tt PolyChord}\footnote{\url{https://github.com/PolyChord/PolyChordLite}.}~\cite{Handley:2015fda,Handley:2015vkr}.

\section{Results and Discussions}\label{sec: results}

Before analyzing the data, we first apply our pipeline to the {\tt Nseries} mocks as a validation test, where our goal is to detect any potential statistical bias arising from our theoretical modelling. In other words, we seek to estimate the theoretical error intrinsic to our modelling choices. For this validation test, we use the mean of the power spectrum multipoles and skew-spectra across the 84 {\tt Nseries} simulations as our data. We compute our covariance matrix using 2048 {\tt Patchy} simulations, the same as the one used for BOSS CMASS NGC. 

We perform two separate analyses using the ${\tt Nseries}$ data. One is to match the effective survey volume of BOSS DR12 ($V_{\rm eff}=6\,h^{-3}\text{Gpc}^3$), which we approximate by rescaling the ${\tt Pathcy}$ covariance matrix by a factor of 2.4. The other one is to match the total volume of the {\tt Nseries} suite ($V_{\rm eff}=235\,h^{-3}\text{Gpc}^3$), this time rescaling the covariance by 84. The two cases serve different purposes. We will use the full {\tt Nseries} volume to estimate potential statistical biases, while using the other one to assess whether it will be significant under the noise level of BOSS DR12. In addition to the $\fnl$ parameters, we vary the biases and counterterms in both analyses. Note that the ${\tt Nseries}$ simulations are only semi-independent, so our estimates for the theoretical error are expected to be conservative.

The result for each effective volume is presented in Table~\ref{tab:nseries}. For each volume, we show constraints for $\fnle$ and $\fnlo$ under three different cases: (1) varying only $\fnle$, (2) varying only $\fnlo$, and (3) varying both $\fnle$ and $\fnlo$. Varying only $\fnle$ we find that the bias on the equilateral shape is around $0.53\sigma$ away from zero when compared with the BOSS level error. However for the orthogonal shape, the theoretical error is at the $4.15\sigma$ level, indicating a significant bias due to our treatment of the survey geometry. Even though in Appendix~\ref{A:validation} we observed that our approximate window convolution leads to an $18\%$ error on the skew-spectra sourced by the orthogonal shape, it is sufficient to cause a severe systematic bias in the full analysis. We arrive at the same conclusion upon jointly varying $\fnle$ and $\fnlo$, finding a significant bias for the latter. As a consequence, in our final analysis on BOSS DR12, we will only vary $\fnle$.

We additionally inspect each skew-spectrum shape to isolate the source of these errors. To be precise, we perform the same ${\tt Nseries}$ validation test as above while restricting the skew-spectra to a single shape $\tildeP_i$ in our data vector. In doing so we find that restricting the data to just $\tildeP_0$ leads to a theoretical error on $\fnlo$ within $0.4\sigma$, relative to BOSS errors. However, the other kernels are more error prone, maintaining a theoretical error of $1-2\sigma$. Our results can be compared with those of Ref.~\cite{DAmico:2022gki}, where the same approximate window convolution is used, albeit on the one-loop bispectrum monopole. They find a negligible theoretical error on $\fnlo$, which is consistent with our checks, since $\tildeP_0$ is nothing more than an average of the bispectrum monopole. These tests suggest that the window-approximation could be used in order to estimate $\fnlo$ using the bispectrum monopole data, but may be too imprecise for the other bispectrum multipoles, at the level of BOSS sensitivity.

\begin{table}[!h]
    \centering
    \renewcommand\arraystretch{1.25}
    \begin{tabular}{c|c|c|c|c}
        Effective Volume ({\tt Nseries}) & \multicolumn{2}{c|}{$V_{\rm eff}=6\,h^{-3}\text{Gpc}^3$} & \multicolumn{2}{c}{$V_{\rm eff}=235\,h^{-3}\text{Gpc}^3$} \\
        \hline\hline
        Parameter varied & $\fnle$ & $\fnlo$ & $\fnle$ & $\fnlo$ \\
        \hline\hline
        $\fnle$ only & $36\pm758$ & \diagbox{}{} & $404\pm81$ & \diagbox{}{} \\
        \hline
        $\fnlo$ only & \diagbox{}{} & $69\pm221$ & \diagbox{}{} & $919\pm24$ \\
        \hline
        $\fnle+\fnlo$ & $241\pm778$ & $56\pm204$ & $194\pm88$ & $827\pm20$ \\
        \hline\hline
    \end{tabular}
    \caption{Posterior means and variances of $\fnle$ and $\fnlo$ with {\tt Nseries} mocks. We consider two different effective volumes, $6\,h^{-3}\text{Gpc}^3$ and $235\,h^{-3}\text{Gpc}^3$, which allows us to determine the size of the systematic errors compared with the noise level of BOSS DR12.}
    \label{tab:nseries}
\end{table}

\begin{figure}[!h]
    \centering
    \includegraphics[scale=0.5]{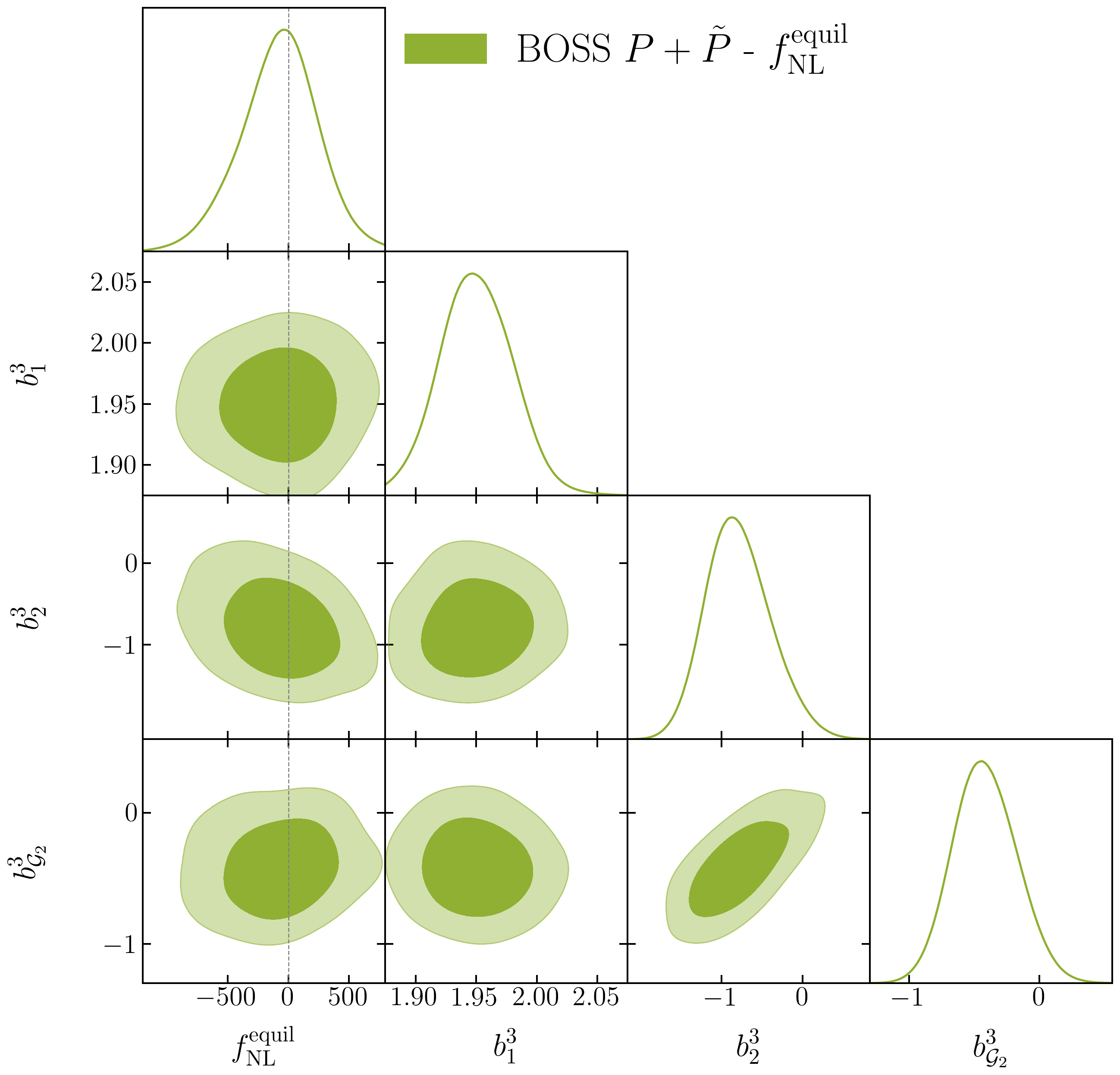}
    \caption{Posterior distributions for our full BOSS analysis, where we vary only $\fnle$ and all the EFT parameters. The bias parameters shown here are for CMASS NGC only, indicated by the index 3. We fix the $\Lambda$CDM cosmological parameters to \textit{Planck} 2018 best-fit values. The expected Poisson shot-noise contribution is subtracted from the BOSS data.}
    \label{fig:MCMCResultBiasfnlEquil}
\end{figure}

Having validated our pipeline and detected any potential bias, we present our main constraint for primordial non-Gaussianity $\fnle$ utilizing the full BOSS DR12 dataset:
\begin{align}
    \text{$\fnle$ varied only:}\quad\fnle=-34^{+296}_{-334}\quad\text{at 68\% CL}\,,
\end{align}
with the contour plot shown in Figure~\ref{fig:MCMCResultBiasfnlEquil}. Our result indicates that $\fnle$ is consistent with zero at 68\% CL even if we include a potential 0.54$\sigma$ statistical bias from our {\tt Nseries} analysis. This result is also consistent within $1\sigma$ with Refs.~\cite{Cabass:2022wjy,DAmico:2022gki}, which find $\fnle=260\pm300$ and $\fnle=245\pm 293$, respectively. As some of the priors on the EFT parameters are poorly known beforehand, we have also checked that widening them by $5\times$ leads to similar constraints for the $\fnle$ parameter.

Let us now compare the results of our main analysis to the ones in the literature. We start with the constraints reported in Ref.~\cite{DAmico:2022gki}, which uses the power spectrum monopole and quadrupole, and the bispectrum monopole, all at the one-loop level. As in our work, Ref.~\cite{DAmico:2022gki} uses the FKP formalism and fixes the $\Lambda$CDM parameters to {\it Planck} 2018 values. However, we use a different set of priors for the EFT parameters. Our choice follows Refs.~\cite{Cabass:2022wjy,Ivanov:2023qzb,Philcox:2021kcw} (discussed in \S\ref{sec: inference}), which employs Gaussian priors with larger variances compared to Ref.~\cite{DAmico:2022gki}. In addition, instead of a lognormal prior on $b_1$ centered at $e^{0.8}$ and Gaussian priors on both $b_2$ and $b_{\G_2}$ centered at zero, we impose uniform priors for all of them.  Overall, we have a comparable variance, i.e. a $7.5\%$ difference between our $\sigma(\fnle)$ and the one of Ref.~\cite{DAmico:2022gki} despite our usage of only the tree-level bispectrum.

We also compare our result with another recent work~\cite{Cabass:2022wjy}, which analyzes the window-less power spectrum multipoles and the bispectrum monopole~\cite{Philcox:2020vbm,Philcox:2021ukg}. In particular, we will focus on the comparison with the aggressive model in Ref.~\cite{Cabass:2022wjy}. This model fixes the $\Lambda$CDM parameters to \textit{Planck} 2018 values. A co-evolution model is also assumed to fix the value of the quadratic biases $b_2$ and $b_{\mathcal{G}_2}$ as a function of $b_1$. Following this strategy, we also see a comparable level of error on $\fnle$, i.e. within 5\% differences on $\sigma(\fnl^{\rm equil})$, between our result and theirs. Overall, our estimation of $\fnle$ is consistent with theirs within $1\sigma$. 

With these comparisons, we can see that even though the skew-spectrum serves as a compression of the full bispectrum, the error bars we present are competitive with the full bispectrum analysis. This was also noted in our previous work~\cite{Chakraborty:2022aok} with harmonic-space skew-spectra in which we have shown, through a Fisher forecast, that skew-spectra with Legendre polynomial kernels are sufficient to constrain not only the bias parameters, but also the seven $\nu\Lambda$CDM parameters. In this work, we include two additional kernels to capture shapes from primordial non-Gaussianity. We see that the addition of these two kernels reduces our error bars on $\fnle$ and the EFT parameters by $10\%-20\%$, compared to the case with just Legendre polynomial kernels.

Finally, we point out some possible sources of systematic errors in our analysis. While our ${\tt Nseries}$ tests reveal an unimportant level of theoretical error on $\fnle$, our posterior mean is still shifted away from zero. Part of these shifts come from our approximate treatment of the window convolution. As mentioned in \S\ref{sec: theory}, this approximation induces up to $5\%$ errors on the  gravitational contribution to the bispectrum, for scales larger than $k=0.03 \, h{\rm Mpc}^{-1}$. In Appendix~\ref{A:validation} we also assess the impact of this approximation on both equilateral and orthogonal shapes, where we found a $6\%$ error for the equilateral shape and $18\%$ error for the orthogonal shape. This error compounds into a large theoretical error for $\fnlo$, as we have seen from the ${\tt Nseries}$ validation tests. Another source of error may be the difference in shot-noise between the {\tt Patchy} and {\tt Nseries} simulations. We tested this by running our pipeline without subtracting the shot-noise from the data and found that our posterior shifts further away by $0.59\sigma$ for $\fnle$, implying that different noise features between the mocks and the data may be a contributing factor. However, we note that the shift from the fiducial $\fnl$ values remains below $1\sigma$. Finally, in our validation test, we have used the ${\tt Patchy}$ mocks in order to approximate the covariance for the ${\tt Nseries}$ suite. This may further be a source of error since the two simulations have slightly different masks.

\section{Conclusions}\label{sec: conclusion}

In this work we present the first analysis of BOSS DR12 galaxy-clustering data using weighted skew-spectra, jointly with the power spectrum multipoles, to probe primordial non-Gaussianity sourced by single-field inflation. Even though the skew-spectra serve the role of a dimension reduction of the full bispectrum, they have been shown to preserve almost all the information for not only amplitude-like parameters, but also $\nu\Lambda$CDM parameters~\cite{MoradinezhadDizgah:2019xun,Chakraborty:2022aok,Hou:2022rcd}. Here we constrain $\fnle$, finding consistency with Refs.~\cite{Cabass:2022wjy} and \cite{DAmico:2022gki}, which use the combination of the galaxy power spectrum multipoles and the bispectrum monopole. 

Motivated by our previous work in harmonic space~\cite{Chakraborty:2022aok}, which shows that the Legendre polynomial kernels for skew-spectra are sufficient to capture most of the information from the full bispectrum, we construct our skew-spectra with five different kernels, including Legendre polynomials, and two additional kernels based on the equilateral and orthogonal primordial bispectrum templates. Each of them can extract non-degenerate information from the galaxy bispectrum. 

Starting from the galaxy catalogue, we apply the FKP formalism to compute both power spectrum multipoles and skew-spectra.  
To model both of these statistics, we follow the effective field theory of large-scale structure approach, which requires several nuisance parameters such as galaxy biases, counterterms and shot-noise parameters. The survey geometry is also taken into account by convolving the targeted spectra with appropriate window functions. In particular we rely on an approximate treatment of the survey geometry for the skew-spectra. 

We assume a Gaussian likelihood and construct our covariance matrix with 2048 {\tt Patchy} mocks. Due to the large number of parameters in our theory model, we use the nested sampling algorithm to estimate the posterior distributions for both $\fnle$ and $\fnlo$, marginalizing over the EFT parameters. We first test our pipeline on {\tt Nseries} mocks and obtain a theoretical error on $\fnle$ to within 1$\sigma$, relative to BOSS level error. However, we find that the window approximation on the primordial bispectrum causes a significant bias on $\fnlo$. Therefore in our final analysis on BOSS DR12 data we choose to only constrain $\fnle$. Fixing the cosmology to {\it Planck} 2018 best-fit values, we obtain $\fnle=-34^{+296}_{-334}$ at 68\% CL, consistent with zero. We find no evidence for primordial non-Gaussianity, which is consistent with the findings of Refs.~\cite{DAmico:2022gki} and~\cite{Cabass:2022wjy}.

This work is a first step in applying the skew-spectra to galaxy survey data. There are some directions in which we plan to make further progress. First, we use an approximate treatment of the window convolution on the bispectrum, which prevents us from reliably estimating $\fnlo$. Moreover, even for the other bispectrum shapes, this approximation restricts the number of large-scale modes accessible to us. Since primordial signals exist in both large-scale and small-scale fluctuations, circumventing this issue would provide a clear path to improving our constraining power. Secondly, we perform a high-dimensional numerical integral in order to obtain our theory skew-spectra, which is a time-consuming task and therefore prevents us from varying the full $\nu\Lambda$CDM parameters simultaneously in the likelihood analysis. This could be potentially accomplished with help of simulation-based inference methods~\cite{Hahn:2022wgo,Hahn:2022zxa,Kobayashi:2021oud,Yuan:2022jqf,Dvorkin:2022pwo}. Lastly, we note that our analysis has used an approximation to determine the normalization of the bispectrum ($I_{33}$ mentioned in \S\ref{sec:galaxyCataogues}), similar to the approximation for $I_{22}$ made at the power spectrum level. This could be a source of systematic error for $\fnl$ (similar to the error on $\sigma_8$ caused by approximating $I_{22}$). While our validation tests verify that this error is small for BOSS data, one should rigorously check this for upcoming surveys.

\acknowledgments

We thank Guido D'Amico, Misha Ivanov, Oliver Philcox, and Georgios Valogiannis for insightful discussions.
CD and SFC are partially supported by the Department of Energy (DOE) Grant No. DE-SC0020223. PC is supported by the DOE Grant No. DE-SC0013607. We acknowledge the use of {\tt CAMB}\footnote{\url{https://camb.info}}~\cite{Lewis:1999bs} and {\tt CLASS-PT}~\cite{Chudaykin:2020aoj} (an extension of {\tt CLASS}~\cite{Blas:2011rf}) for theoretical computation; {\tt pycuba}~\cite{Hahn:2004fe,Buchner:2014nha} for numerical integration; {\tt nbodykit}~\cite{Hand:2017pqn} and {\tt Corrfunc}~\cite{Sinha:2019reo} for analysis of simulations and data; {\tt PolyChord}~\cite{Handley:2015fda,Handley:2015vkr} for parameter inference, and {\tt GetDist}\footnote{\url{https://getdist.readthedocs.io/en/latest}}~\cite{Lewis:2019xzd} to generate posterior distributions.
\clearpage

\appendix

\section{BAO Damping}\label{A:IR}

The damping of the BAO feature at low redshifts can be captured by performing an infrared resummation on the so-called wiggly part of the matter power spectrum \cite{Ivanov:2018gjr}. For this, we first need to split the wiggly and non-wiggly parts of the matter power spectrum. Instead of using the standard Hu-Eisenstein fitting formula \cite{Eisenstein:1997jh} for the smooth part of the power spectrum, we isolate the BAO feature, following Refs. \cite{Hamann:2010pw,Chudaykin:2020aoj}. 
This procedure uses the Discrete Fast Sine transformation (DST) to identify the BAO bump in the real-space correlation function, and thus separate it to obtain the so-called wiggly part of the power spectrum. 

Once we get the wiggly ($P_{\rm w}$) and non-wiggly ($P_{\rm nw}$) parts of the matter power spectrum, we can obtain the infrared resummed power spectrum:
\begin{align}
    P_{\rm L, IR}(k) &= P_{\rm nw}(k) + e^{-k^2\Sigma_{\rm tot}^2}P_{\rm w}(k)\,.
\end{align}
Here $\Sigma_{\rm tot}^2$ is defined as \cite{Philcox:2022frc, Ivanov:2018gjr}:
\begin{align}
    \Sigma^2 &= \frac{1}{6\pi^2}\int_0^\Lambda \d q P_{\rm nw}(q)\left[1-j_0(q x_{\rm osc})+2j_2(q x_{\rm osc})\right],\nonumber\\
    \Sigma_{\rm tot}^2 &= -\frac{2}{15}f^2\delta\Sigma^2 + \left(1+\frac{1}{3}f(2+f)\right)\Sigma^2,\nonumber\\
    \delta\Sigma^2 &= \frac{1}{2\pi^2}\int_0^\Lambda \d q P_{\rm nw}(q)j_2(q x_{\rm osc}),
\end{align}
and we set $\Lambda=0.1$ $h$Mpc$^{-1}$ and $x_{\rm osc}=110$ $h^{-1}$Mpc.

\section{Survey Geometry Approximation}\label{A:validation}

As mentioned in $\S$\ref{sec:surveyGeometry}, we approximate the impact of the survey geometry in our theoretical skew-spectra computation following Refs.~\cite{Gil-Marin:2014sta,Gil-Marin:2016wya} where we only convolve factors of the power spectrum with the window function. Since we are working with non-Gaussian initial conditions, our bispectrum, and therefore skew-spectra, contain both contributions due to gravitational nonlinearities and also due to the primordial bispectrum. References~\cite{Gil-Marin:2014sta,Gil-Marin:2016wya} only assess this approximation on the gravitational terms, finding at most an error of $5\%$. In this appendix, we will quantify the error on our skew-spectra due to this approximation, specifically targeting the primordial contribution.

As a first step, we estimate the exact window convolved skew-spectra. To do so we will generate a large number of synthetic non-Gaussian density fields, appropriately masked. From each of these we may estimate our desired skew-spectrum, ultimately averaging over all of our mocks. To generate our boxes we use the same survey specification as the CMASS NGC patch (see $\S$\ref{sec: data}).

In order to generate our synthetic density field, we start with a Gaussian random field $\phi$ generated using the primordial power spectrum $P_{\phi}(k)=A_s k^{-3}$ \footnote{For the purpose of this test we set $n_s=1$ in order to simplify the analysis. We do not expect this choice to significantly impact our conclusions.}. Next, we generate a non-Gaussian field $\Phi$ for both equilateral and orthogonal configurations from this Gaussian random field $\phi$ as~\cite{Scoccimarro:2011pz,Coulton:2022qbc}:
\begin{align}
    \Phi_{\rm equil} &= \phi + \fnl\left[-3\phi^2 + 4\partial^{-1}(\phi\partial\phi) + 2\nabla^{-2}(\phi\nabla^2\phi)-2\nabla^{-2}(\partial\phi)^2\right]\\
    \Phi_{\rm ortho} &= \phi + 3\fnl\left[-\left(1+\frac{9p}{27}\right)\phi^2-\left(2+\frac{60p}{27}\right)\frac{1}{3}\nabla^{-2}(\partial\phi)^2\right.\nonumber\\
    &\left.+2\left(1+\frac{15p}{27}\right)\left((1-t)\partial^{-1}(\phi\partial\phi)+t\nabla^{-2}(\phi\nabla^2\phi)\right)\right.\nonumber\\
    &\left.+\frac{p}{27}\nabla^2(\partial^{-1}\phi\partial^{-1}\phi)-\frac{20p}{27}\partial^{-1}(\nabla^2\phi\partial^{-1}\phi)-\frac{12p}{27}\partial(\phi\partial^{-1}\phi)+\frac{30p}{27}\partial\phi\partial^{-1}\phi\right]\,,
\end{align}
where $t=\tfrac{2+60p/27}{6\left(1+15p/27\right)}$ and $p=8.52587$. Note that $\Phi_{\rm equil}$ will have a nontrivial bispectrum equal to the equilateral shape, and likewise for $\Phi_{\rm ortho}$. These parametrizations also ensure that the largest correction to the power spectrum scales as $k^{-2}$. We subsequently evolve this primordial density field into a late-time matter density field by multiplying by the matter transfer function $T(k,z)$. In order to simulate a galaxy overdensity field, we include a factor of $b_1+f\mu^2$ to simulate the galaxy bias $b_1$ and redshift-space distortion. To account for the survey geometry, we directly apply the meshed FKP weight $1/(1+\bar{n}_g(z)P_{\rm FKP})$ on each pixel in our simulation box.

With this pipeline for generating the non-Gaussian field, we simulate $10,000$ such maps for both equilateral and orthogonal PNG. Note that by construction, these synthetic fields only contain non-Gaussian information due to the primordial contribution. This enables us to isolate the contribution to each of the skew-spectra sourced by PNG. We then average over skew-spectra estimated from each of these boxes in order to obtain an estimate of the exact window convolved skew-spectra.

In Table~\ref{tab:skewValidation} we list the percentage difference between the exact window convolved skew-spectra and those computed using the approximation specified in $\S$\ref{sec:surveyGeometry}. Note that in quoting these numbers, we discard $k$ modes for which the skew-spectra almost vanish, since these modes are noisier and in any case contain too little power to impact $\fnl$ constraints. We see that for equilateral PNG, the fractional errors are well within at most $6.3\%$ except for $\tilde{P}_{\rm ortho}$. However, this is not problematic since this shape has limited sensitivity to $\fnle$. On the other hand, for orthogonal PNG, the fractional errors are in general larger, ranging from $9.2\%$ to $18.3\%$.

\begin{table}[h!]
    \centering
    \renewcommand\arraystretch{1.25}
    \begin{tabular}{c|c|c|c|c|c}
        \hline\hline
        Skew-spectrum & $\tilde{P}_0$ & $\tilde{P}_1$ & $\tilde{P}_2$ & $\tilde{P}_{\rm equil}$ & $\tilde{P}_{\rm ortho}$ \\
        \hline\hline
        Equilateral PNG & 4.6\% & 6.3\% & 5.3\% & 3.6\% & 15.4\%\\
        \hline
        Orthogonal PNG & 18.3\% & 10.0\% & 14.8\% & 17.8\% & 9.2\%\\
        \hline\hline
    \end{tabular}
    \caption{Fractional error between the exact window convolved skew-spectra (obtained from the average over simulations) and the approximate theoretical computation. We consider both equilateral and orthogonal configurations, and for each skew-spectrum we show the maximum percent error across all $k$ modes.}
    \label{tab:skewValidation}
\end{table}

\bibliographystyle{utphys}
\bibliography{refs.bib}
\nocite{*}

\end{document}